\documentclass[manuscript,screen]{acmart}

\usepackage{tikz}
\usepackage{tikzsymbols}
\usetikzlibrary{positioning}

\usepackage{mdframed}
\mdfsetup{skipabove=5pt,skipbelow=5pt}
\usepackage{enumitem}

\usepackage{longtable}

\usepackage{caption}
\usepackage{subcaption}

\usepackage[T1]{fontenc}
\usepackage[utf8]{inputenc}

\AtBeginDocument{%
  \providecommand\BibTeX{{%
    \normalfont B\kern-0.5em{\scshape i\kern-0.25em b}\kern-0.8em\TeX}}}

\setcopyright{acmcopyright}
\copyrightyear{2023} 
\acmYear{2023} 
\setcopyright{rightsretained}

\acmConference[GoodIT '23]{ACM International Conference on Information Technology for Social Good}{September 6--8, 2023}{Lisbon, Portugal}
\acmBooktitle{ACM International Conference on Information Technology for Social Good (GoodIT '23), September 6--8, 2023, Lisbon, Portugal}

\acmDOI{10.1145/3582515.3609573}
\acmISBN{979-8-4007-0116-0/23/09}

\usepackage[nameinlink]{cleveref}

\usepackage{multirow}

\usepackage[obeyFinal]{easy-todo}

\begin{document}

\title{What Twitter Data Tell Us about the Future?}


\author{Alina Landowska}
\authornotemark[1]
\affiliation{%
  \institution{SWPS University of Social Sciences and Humanities}
  \city{Warsaw}
  \country{Poland}
}
\email{alandowska@swps.edu.pl}
\orcid{0000-0002-7966-8243}

\author{Marek Robak}
\authornotemark[2]
\affiliation{%
  \institution{Cardinal Stefan Wyszyński University in Warsaw}
  \city{Warsaw}
  \country{Poland}
}
\email{m.robak@uksw.edu.pl}
\orcid{0000-0001-8331-2891}

\author{Maciej Skórski}
\authornotemark[3]
\affiliation{%
  \institution{University of Warsaw}
  \city{Warsaw}
  \country{Poland}
}
\email{maciej.skorski@gmail.com}
\orcid{0000-0003-2997-7539}

\renewcommand{\shortauthors}{Landowska, Robak, and Skórski}

\begin{abstract}
Anticipation is a fundamental human cognitive ability that involves thinking about and living towards the future. While language markers reflect anticipatory thinking, research on anticipation from the perspective of natural language processing is limited. This study aims to investigate the futures projected by futurists on Twitter and explore the impact of language cues on anticipatory thinking among social media users. We address the research questions of what futures Twitter's futurists anticipate and share, and how these anticipated futures can be modeled from social data. To investigate this, we review related works on anticipation, discuss the influence of language markers and prestigious individuals on anticipatory thinking, and present a taxonomy system categorizing futures into "present futures" and "future present." This research presents a compiled dataset of over 1 million publicly shared tweets by future influencers and develops a scalable NLP pipeline using SOTA models. The study identifies 15 topics from the LDA approach and 100 distinct topics from the BERTopic approach within the futurists' tweets. These findings contribute to the research on topic modelling and provide insights into the futures anticipated by Twitter's futurists. The research demonstrates the futurists’ language cues signals futures-in-the-making that enhance social media users to anticipate their own scenarios and respond to them in present. The fully open-sourced dataset, interactive analysis, and reproducable source code are available for further exploration.
\end{abstract}




\received{10 June 2023}

\maketitle

\section{Introduction}

\subsection{Background}
\emph{What is anticipation about?} Anticipation is the human cognitive ability of thinking about and living towards the future. It is a natural part of our lives and may be found in almost any situation. Anticipatory thinking and behaving are necessary to manage a complex and unforeseeable future ~\cite{poli_complexity_2009}. A function of the human brain as “anticipation machine” is to prepare people for future events with a whole range of probabilities (from low to high), not simply predict what might happen ~\cite{gilbert_prospection_2007}. Most of the time, people are engaged in an effort to figure out how to predict particular occurrences. In the face of complexity and uncertainty generated by an ever-changing world, anticipation plays a crucial role in managing the challenges and opportunities. Understanding how individuals anticipate the future is vital for effective future management.

Language markers reflect anticipatory thinking ~\cite{bennett_anticipation_1976}. However, research on anticipation from the perspective of natural language processing has yet to be conducted. In this study, we aim to investigate the futures projected by futurists on Twitter, assuming that social media users anticipate their futures based on language cues provided by these futurists. Specifically, we seek to answer the following research questions: 
\todo{can we refer to these specific questions?}
\begin{mdframed}
\begin{enumerate}[label=\textbf{RQ.\arabic*}]
\item \label{RQ.1} What futures do Twitter's futurists anticipate and share?
\item \label{RQ.2} How can anticipated futures be modelled from social data?
\end{enumerate}
\end{mdframed}
We discuss the development of anticipation research through different concepts, the features of anticipatory thinking, and approaches to categorizing futures. By reviewing related works, we establish a foundation for our investigation into Twitter's futurists and their projected futures. 

Our research aims to shed light on the futures anticipated by Twitter's futures influencers and explore the impact of language cues on anticipatory thinking among social media users. Through the analysis of social data, specifically futurist's tweets, and employing the LDA and BERT models, our aim is to deepen our comprehension of the anticipated futures that emerge within the digital realm. This study contributes to the growing body of research on anticipatory rhetoric and its implications for anticipation thinking.

\Cref{sec:anticipation_language} discusses the influence of language markers and prestigious individuals, like futurists, on others' anticipatory thinking, providing the foundation for our research. In Section 2.2,  we present a taxonomy system that categorizes futures based on their relation to the past and present, setting the stage for understanding the futures projected by Twitter's futurists. Section 3 describes the data acquisition and pre-processing (3.1), modelling (3.2), and inference (3.3). In Section 4, we present and discuss our results obtained from the LDA and the BERT modelling.

\subsection{Contribution}

Our technical contributions are:
\begin{itemize}
\item \textbf{Future Influencers Dataset} we contribute a compiled dataset of more than 1M tweets shared publicly by Future Influencers. We believe that this interesting  dataset will be useful for other research projects.
\item \textbf{Open-Sourced \& Post-API} analysis, including source code, data and several interactive presentations, is available as the Open Science Framework repository~\cite{skorski_what_2023}. Given \href{https://developer.twitter.com/en/products/twitter-api/academic-research}{limitations of Twitter's Academic License}, as well as general concerns around data availability in the Post-API world~\cite{tromble_where_2021} we decided to scrape the publicly available data, and demonstrated that this is both accurate and efficient.

\item \textbf{Scalable NLP pipeline with SOTA models}: we leverage established topic extraction methods and implement an efficient pipeline, tweaked to Twitter data. The code is available in the repository~\cite{skorski_what_2023}. 
\end{itemize}

\section{Related Works}

In this section, we discuss the foundations of anticipation (Section 2.1) and group related works into concepts contributing to the features of anticipatory thinking (Section 2.2).

\subsection{Anticipatory cues in language}\label{sec:anticipation_language}

\emph{Why do people anticipate?} Human's ability to combine knowledge, experiences, stories, or concepts and formulate certain conclusions for the futures allows us to deal with uncertainty and ambiguity ~\cite{klein_anticipatory_2011}. Knowing if  the future event will occur (i.e., the hazard rate, HR) and when it will happen (i.e., time estimation) are two key factors determining this uncertainty ~\cite{grabenhorst_two_2021}. Anticipation helps people manage the unknown future. On one side, it is a forward-thinking mindset ~\cite{poli_handbook_2019}. On the other, it is the application of the former's result to action ~\cite{poli_handbook_2019}. Research shows that there is a human tendency to anticipate positive future events over negative ones, e.g., 
\cite{bennett_anticipation_1976, grabenhorst_anticipation_2019}.

\emph{How are people attracted to anticipated futures?} Humans’ attention is primarily attracted by the mental projection of a scenario onto a possible future ~\cite{moore_data-frame_2011}. While attention is focused on certain events (or information), ones are intentionally neglected; when sensory cues (e.g., visual cues) are signalled, then human anticipation is activated ~\cite{grabenhorst_anticipation_2019}. Attracting attention with content plays a crucial role on social media platforms. The content, containing some anticipated futures, might particularly attract the attention of the followers. By paraphrasing Rosen’s model ~\cite{rosen_foundations_1972}, it can be said that a number of followers might anticipate the future on the basis of the input information (i.e., social media content in a form of posts/tweets/videos about the future state of, e.g., future events, the economy, health system, etc.) gained from some prestigious individual.

\emph{Who is signalling cues that activate anticipatory thinking?} The Foggian approach states that persuasive technologies, e.g., social media platforms such as Twitter) aim at changing people’s attitudes, which in turn modify behaviours ~\cite{fogg_persuasive_2003, holmes_ethos_2016}. One of Twitter’s persuasive elements is influencers using their privileged position; when someone is prestigious, it means the person is honoured and esteemed, and people tend to listen to them ~\cite{henrich_evolution_2001}. Prestigious status is given to a person who shares knowledge, assets, skills, and/or other resources with other people ~\cite{boyd_origin_2005, henrich_evolution_2001, patton_reciprocal_2000, pinker_how_1997, tooby_friendship_1996}. Because people are "default infocopiers", they attempt to learn directly from prestigious individuals rather than “reinvent the wheel” which leads to significant cost savings ~\cite{henrich_evolution_1998, henrich_evolution_2001}. Adopting this approach, we recognize 'future influencers', who at the same time are prestigious individual and futurists, as those who impact others' anticipatory thinking by tweeting anticipatory cues in language markers. As research proved, the credibility and trustworthiness of such a future influencer determine the success of the content spread, e.g., ~\cite{jin_following_2014, spry_celebrity_2011}.

\emph{How do language markers serve as carriers for the sharing of anticipated futures?} The development of language made possible, among other things, the social transfer of a significant amount of information that is impossible to deduce based just on auditory and visual senses e.g., ~\cite{henrich_evolution_2001}. On social media platforms, such as Twitter, people with a prestige status share the information with others ~\cite{acerbi_cultural_2019, acerbi_cultural_2016}. Many different types of information are shared, e.g., ~\cite{acerbi_storytelling_2022, berger_contagious_2013, gloor_identifying_2019}). Language markers in the form of written words are recognized as visual cues that reflect anticipatory thinking ~\cite{bennett_anticipation_1976}. Tweets attract others attention, especially like-minded others \cite{mosleh_cognitive_2021}, and this is their primary function. Further, they might contain anticipatory cues coded in language for social media users, which might guide their own anticipation about 'what' and 'when' \cite{momennejad_human_2012}.

\subsection{Futures Taxonomy}

\emph{What are the types of futures?} For the purpose of this research, we recognize two main categories of the future: a. ‘present futures’ ~\cite{luhmann_differentiation_1982} are ‘pre-given futures’ rooted in the past, i.e., lived ones ~\cite{adam_futures_2011}; and b. ‘future present’ ~\cite{luhmann_differentiation_1982} are ‘futures in the making’ that are possibly latent, growing, and changing ones, i.e., living ones ~\cite{adam_futures_2011}. Following Poli‘s ~\cite{poli_introduction_2017, poli_anticipation_2014} taxonomy, the present future refers to the concrete (practical, presentational, explicit) and it is the contextual, embodied, and embedded future ~\cite{adam_future_2007}; and the future present refers to the abstract (symbolic, representational, implicit) future that is the “decontextualised future emptied of content, which is open to exploration and exploitation, calculation, and control” ~\cite{adam_future_2007, poli_anticipation_2014, poli_introduction_2017}.

Present futures are linear continuations of the past in the present ~\cite{poli_introduction_2017}, which shape the future by present means ~\cite{miller_futures_2007}. Shaping by imagination, planning, projection, and production is a need of the present that looks for economic or scientific forecasts ~\cite{adam_future_2007}. These concrete-like futures are closed-type futures that can be calculated. Hence, present futures look as one's viewpoint \emph{versus} another's viewpoint ~\cite{luhmann_differentiation_1982}. The value of the present future is calculated against its alternatives and traded as a commodity; On the basis of its expected returns, one future outcome may be traded for another ~\cite{adam_future_2007}, and the future with the highest value is the one with the largest profit. Because of the optimistic or pessimistic character of their transformed or reinterpreted visions, the present futures serve as utopian projections aiming at evoking hope or fear ~\cite{luhmann_differentiation_1982}. Despite that, they seem to be the most realistic ones because they are common sense-based from the present perspective. 

Future presents are latent, but they can be recognized and foreseen ~\cite{adam_futures_2011} thus impacting the present by entering into it and being used in the present. These futures are conceptualized, isolated from their contexts, and technologically prejudiced ~\cite{luhmann_differentiation_1982}, which makes them limitless possibilities, leading to high uncertainty ~\cite{adam_future_2007, beckert_capitalism_2013,poli_anticipation_2014,poli_introduction_2017}. These abstract futures are open-type futures with abstract values that are freely traded on the assumption that they can be computed at any location and at any time, and it may be used for any situation ~\cite{adam_future_2007}. Like fantasies, future presents are unpredictable and evoke fictional expectations based on the “as if” rule ~\cite{beckert_capitalism_2013}. However, they emerge from abstract systems as a result of scientific modelling based on past data (e.g., forecasting extrapolation) ~\cite{giddens_modernity_1991}, and they are quantitative by nature, so they might be ranked and compared to one another ~\cite{poli_introduction_2017}. Future presents are anticipatory in their character because of  their ‘use-of-the-future’ orientation ~\cite{poli_introduction_2017}.


\section{Research Methods}

\begin{figure}
    \centering
    \resizebox{0.5\linewidth}{!}{
    \includegraphics[width=0.99\linewidth]{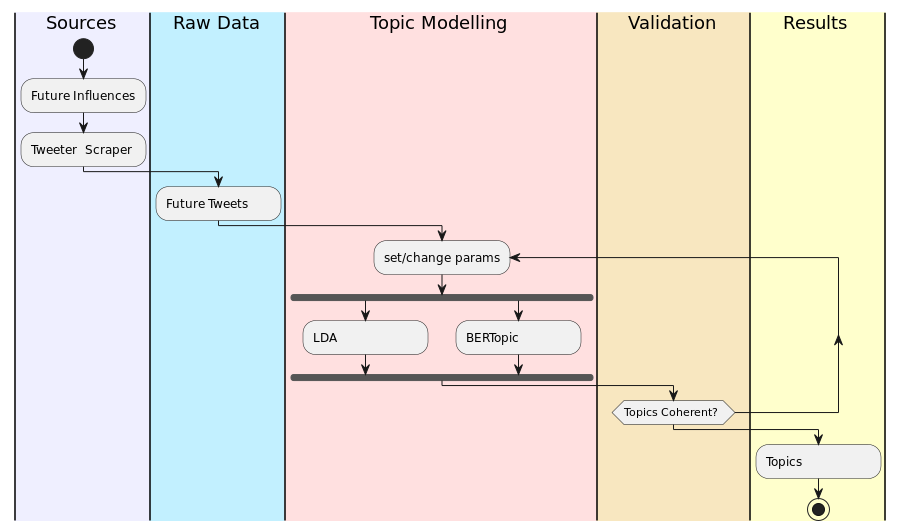}
    }
    \caption{Our workflow: from data acquisition, through  topic modelling with validation feedback, until topics inference.}
    \label{fig:workflow}
\end{figure}

In this section, we describe in detail our workflow.
There are three main stages: data collection, topic modelling (followed by validation feedback), and inference (classification with visualization). The process is illustrated on \Cref{fig:workflow}.



\subsection{Data Acquisition}

The data set was obtained using the scraping library \texttt{snsscrape}~\cite{justanotherarchivist_snscrape_2023}, and Tweets were gathered between the 1sth of January 2021, and 31th of March 2023. There are 1,2 milions tweets in the 'future' data set. The data set comprises tweets posted by 'future influencers' that were carefully selected by human experts. The final list included about 250 futurists who were expected to make references to their anticipated future scenarios in their tweets.



Here our contribution is the successful use of the scrapping API~\cite{justanotherarchivist_snscrape_2023} at large scale, demonstrating that a fully open-source data-mining on Twitter is possible in the Post-API world.



\subsection{Topic Modelling}

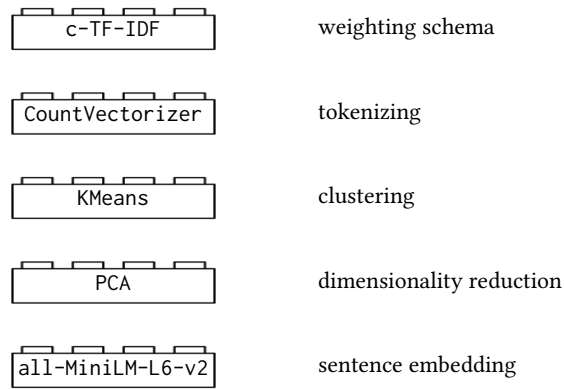
\begin{figure}[!h]
\pgfdeclareimage[width=3cm, height=1.75cm]{lego}{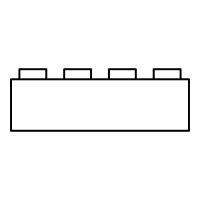}
\begin{tikzpicture}
\node (embed) [label=center:\texttt{all-MiniLM-L6-v2}] {\pgfuseimage{lego}};
\node [right=1 cm of embed] {sentence embedding};
\node (reduce) [above=-0.85cm of embed, label=center:\texttt{PCA}] {\pgfuseimage{lego}};
\node [right=1 cm of reduce] {dimensionality reduction};
\node (cluster) [above=-0.85cm of reduce, label=center:\texttt{KMeans}] {\pgfuseimage{lego}};
\node [right=1 cm of cluster] {clustering};
\node (countvect) [above=-0.85cm of cluster, label=center:\texttt{CountVectorizer}] {\pgfuseimage{lego}};
\node [right=1 cm of countvect] {tokenizing};
\node (weight) [above=-0.85cm of countvect, 
label=center:\texttt{c-TF-IDF}] {\pgfuseimage{lego}};
\node [right=1 cm of weight] {weighting schema};
\end{tikzpicture}
\caption{The architecture of the BERTopic pipeline used in this work.}
\label{fig:bertopc_arch}
\end{figure}

Following prior work on extracting topics from Twitter data~\cite{bogdanowicz_dynamic_2022,meddeb_using_2022,egger_topic_2022}, 
we trained two state-of-the-art models: the Latent Dirichlet Allocation model (LDA) provided by the \texttt{Gensim} library~\cite{r_rehr_uv_rek_software_2010} (we used the multi-core implementation), and the BERTopic model implemented in the \texttt{BERTopic} library, with submodels (embeddings, dimensionality reduction, clustering, tokenizing and class-weighting) chosen to support batch training, as detailed in \Cref{fig:bertopc_arch}. Note that we couldn't use some more accurate algorithms (such as Uniform Manifold Approximation and Projection for dimensionality reduction) if their implementations did not support online training.

\subsection{Inference}

Hyperparameters of both models were optimized  by considering the coherence measure and expert judgment\todo{optimization packages like \cite{terragni_octis_2021}}. The heaviest part 
of the BERT model - the transformer layer - was optimized by precomputing sentence transformers on GPU. In turn, the heaviest part of the LDA model - text preprocessing - was optimized by parallelizing.
Both models were trained on batches (online) to ensure scalability, so that the memory consumption was constant with respect to the data size. The LDA model was trained on multiple CPU cores.

\section{Results and Discussion}

In this section, we present what we have accomplished so far in terms of futures' topic modelling by using the LDA and the BERT approaches, as well as discuss these findings in terms of futures as described in Section 4, trying to recognize futures that are anticipated by 'future influencers'. 
Due to the space constraints, we refer readers to the repository~\cite{skorski_what_2023} for details and interactive content.

\subsection{Futures from the LDA Topic Modelling}

\begin{figure}[h!]
\centering
\begin{subfigure}{0.53\textwidth}
\centering
\includegraphics[height=4cm]{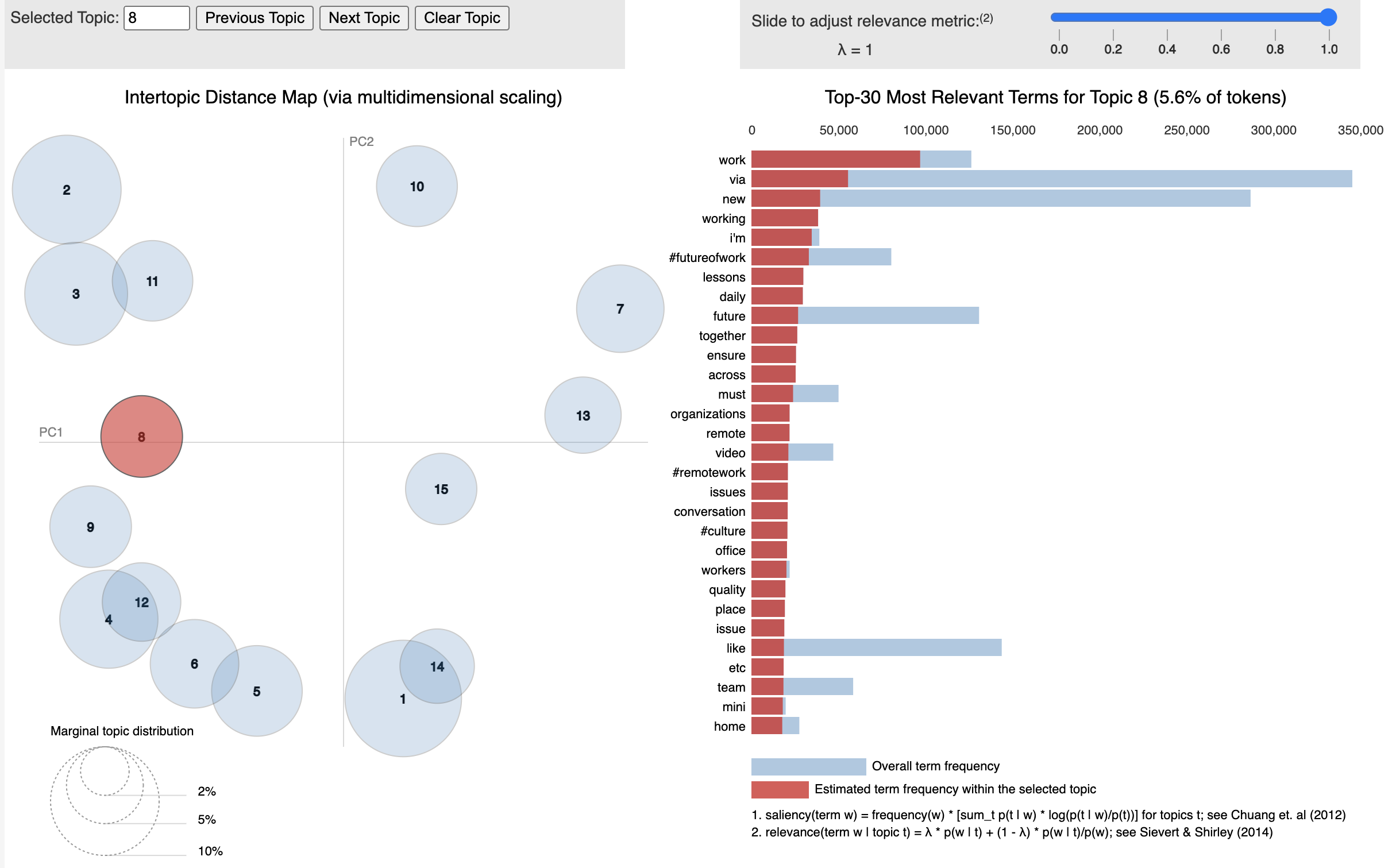} 
\caption{LDA topics. The cluster in red is the "future of work" topic, with keywords such as "\#remotework", "culture", "office", "conversation".}
\end{subfigure}
\hfill
\begin{subfigure}{0.45\textwidth}
\centering
\includegraphics[height=4cm]{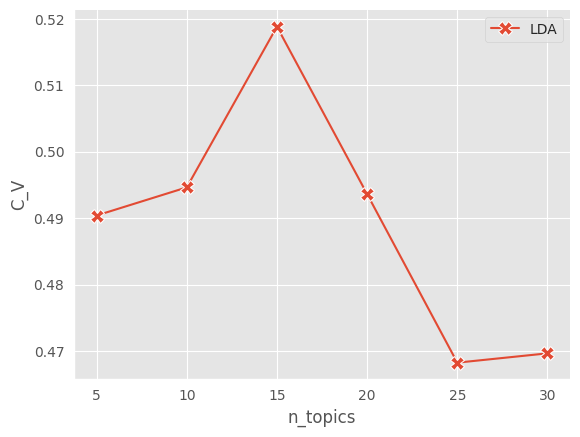}
\caption{Validation with the $C_V$ coherence score suggests $n=15$ as the optimal number of topics.}
\end{subfigure}
\caption{Validation of the trained LDA model.}
\label{fig:lda_validation}
\end{figure}

By applying the LDA method and performing additional validation with the established coherence score developed by Röder et al.~\cite{roder_exploring_2015} (see Figure~\ref{fig:lda_validation} and the repository for more details), 15 topics were identified in the data set. The boundaries of topics are not always clearly defined, so topics overlap with each other. For instance, a topic dedicated to the future of AI, which includes i.a., machine learning, learning, data science, big data, coding, and analytics, overlaps with a topic dedicated to the future of technology, which includes i.a., cybersecurity, metaverse, technology, IoT, behaviour, leadership, management, and startups. Because the tweets are very short text utterances (a regular tweet is limited to 280 characters) and there might be too many topics within the analysed dataset, it clearly becomes unsuitable for the LDA. Overall, the model achieved a decent (in its class) coherence score of more than 50\%. 

\begin{figure}
\centering
\includegraphics[width=0.99\textwidth]{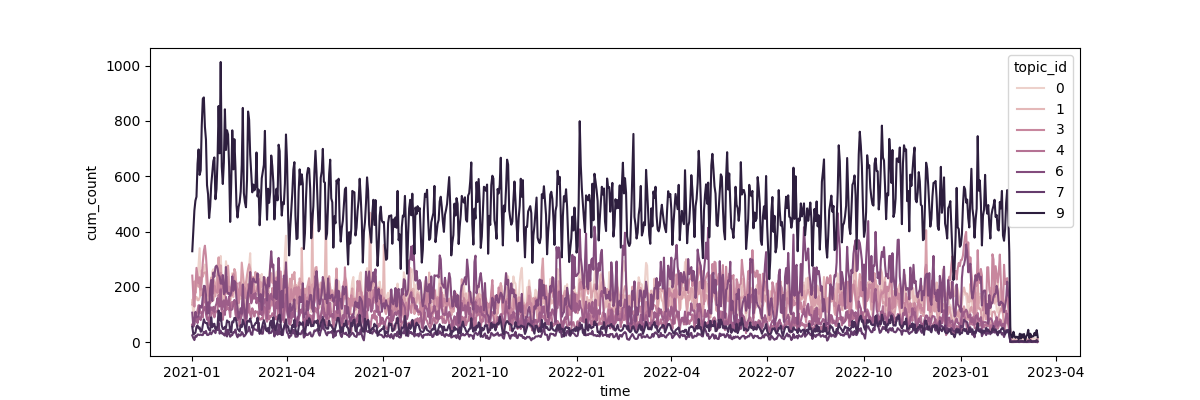}
\caption{Topics dynamic in time.}
\label{fig:topics_dynamic}
\end{figure}
As we see on \Cref{fig:topics_dynamic} all topic contributions remain consistent in time. It suggests that the themes or subjects covered by the topics continue to be relevant and maintain their significance without evident fluctuations or shifts in prominence. This consistency implies that the topics consistently capture and represent the main content on future within the given dataset over time.  

\subsection{Futures from the BERT Topic Modelling}

The sentence transformers applied in BERTopic allow for the extraction of embeddings that capture the semantic meaning of sentences. This enables the clustering of documents based on not only lexical but also semantic similarities. In turn, the c-TF-IDF algorithm, used to generate dense clusters, enables us to visualize a readily understandable list of topics while retaining significant or defining words in the topic description. Hence, the implementation of the BERTopic enabled the extraction of significantly more extensive and detailed topics than the LDA. The results revealed discussions, which were not identified in the LDA, for instance anticipated trends in robotics or expectation on COVID-19 research progress. Finally, the model proposed a total of about 100 hierarchically organized clusters of good coherence (65\%), that can be further merged to about 20 well-separated topics (albeit with a loss in coherence). Among the merged topics are: future of AI, automation, digital transformation, future of work, expected changes in management culture, privacy concerns around social media,  These potential futures are conceptualized and detached from their specific contexts. Consequently, they represent boundless possibilities and must force social media users reading them to anticipate their own futures, leading to a state of high uncertainty. For instance, the future of automation is expected to generate approximately 133 million new job opportunities, despite an estimated loss of 75 million jobs. However, the overall employment rate is expected to continue its upward trajectory, so people reading this might anticipate a skill shift and, in consequence, a need for upskilling or even reskilling.

\begin{figure}[h!]
\centering
\begin{subfigure}{0.48\textwidth}
\centering
\includegraphics[height=10.5cm]{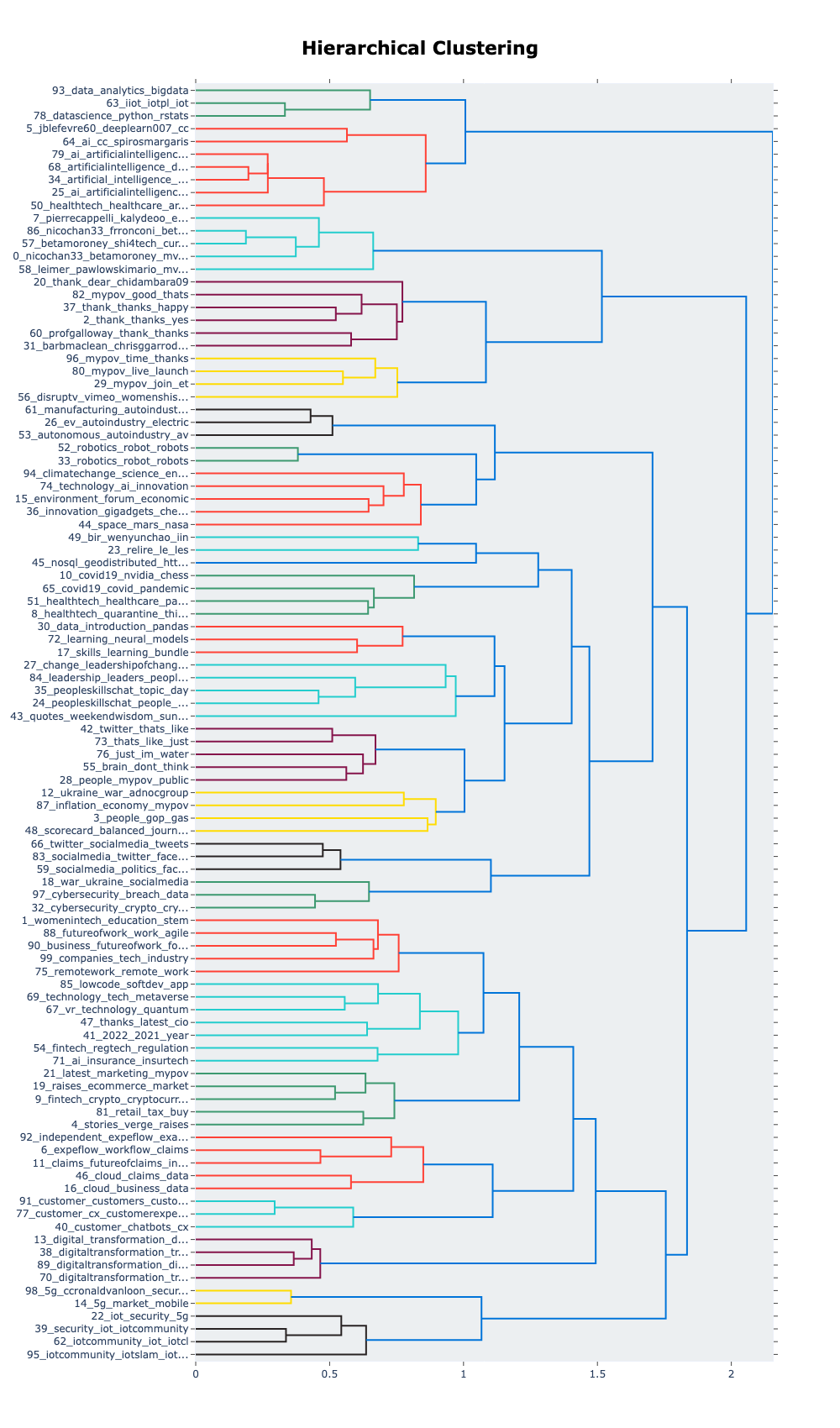}
\caption{The hierarchy of topics identified by the BERTopic model.}
\label{fig:bertopic_hierarchy}
\end{subfigure}
\
\begin{subfigure}{0.48\textwidth}
\centering
\includegraphics[width=0.99\textwidth]{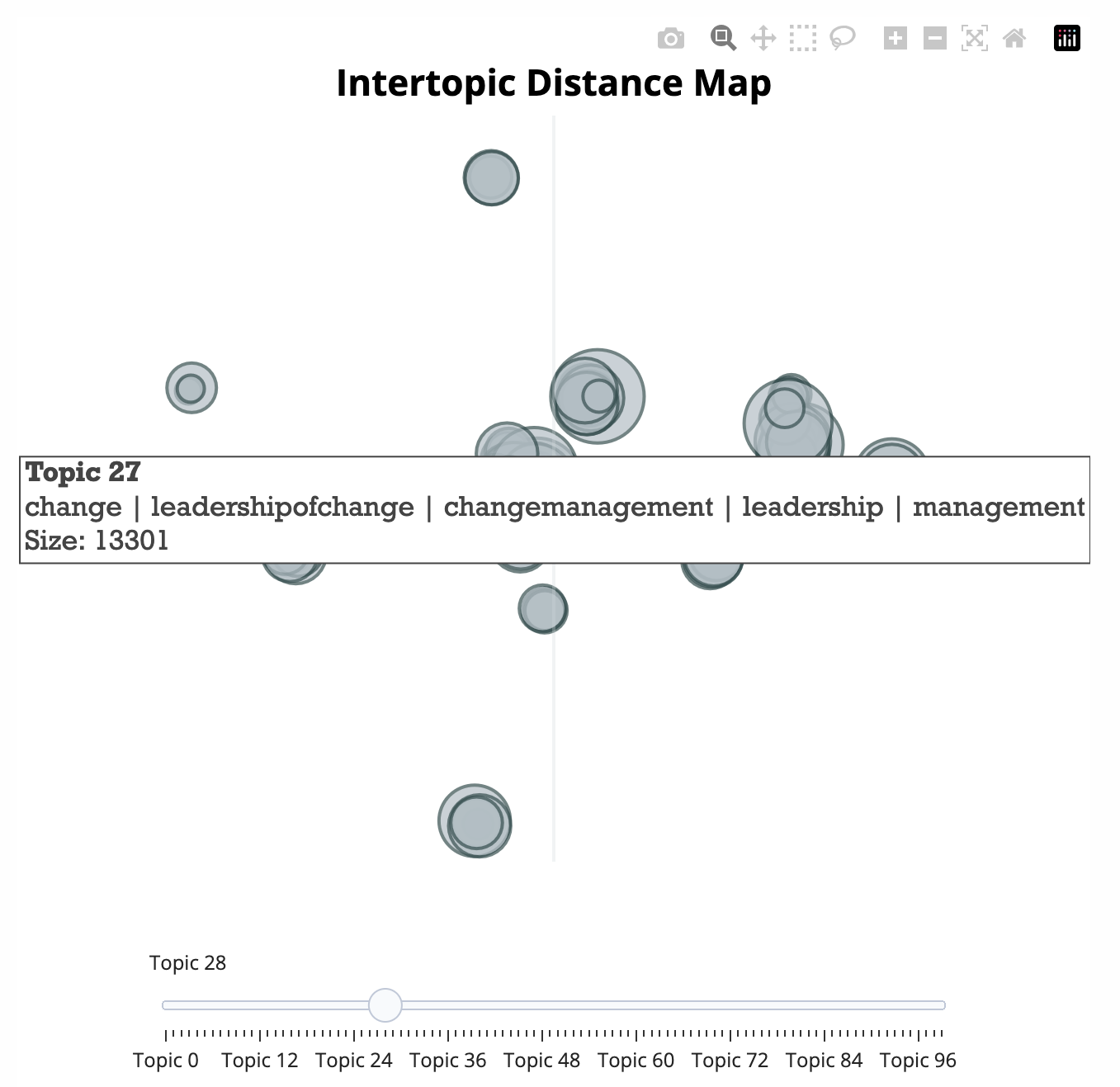}
\caption{The inter-topic distance map (before topics reduction).}
\label{fig:bertopic_distance}
\end{subfigure}
\caption{Results from BERTopic.}
\end{figure}

\subsection{Limitations}

While our study provides valuable insights, it is important to acknowledge its limitations. Firstly, the analysis was confined to a specific subset of Twitter users (namely, influencers), which may slightly restrict the generalizability of our findings. Therefore, for future research we plan to investigate the generalizability of our findings to broader and more representative future influencer populations, specifically including those who were not included in the analysis.
Another limitation is that not all tweets attributed to future influencers necessarily reflect their own future anticipations. Some of these tweets could have been retweeted from other users or may contain quotations. 

Additionally, the study was limited to tweets published by future influencers. It would be worthwhile to extend the research to include user comments and examine the sentiment associated with those comments—essentially, exploring how users perceive the future suggested by future influencers.


\section{Conclusion}
This study examines the topics referring to the futures introduced by futurists on the widely-used social media platform, Twitter. Twitter was selected as the platform of choice due to its global reach and its ability to empower influencers, including futurists, to express their viewpoints on upcoming events, technologies, and processes that they anticipate, and communicate through tweets.
In this paper, the LDA and the BERTopic models, were used to identify the most frequently discussed topics in the dataset of 'future influencers' tweets, and the results were compared between both models. Firstly, our process began with a thorough data cleansing phase aimed at eliminating any irrelevant elements. Subsequently, we employed the LDA and the BERTopic on the pre-processed data. In comparison to the LDA, the BERTopic is more flexible and gives more significant and varied results. 

It is evident that the futures anticipated by future influencers and shared by them in tweets are the future present types that are not fixed entities; instead they represent dynamic and evolving states that undergo constant shaping and transformation, e.g. the future of digital twins holds latent potential that is yet to be fully realized - there may be untapped capabilities, unexplored use cases, and advancements that can transform how digital twins are utilized and integrated into various industries and domains. These 'futures in the making' are continuously developing, and subject to change, indicating their vibrant and alive nature. Because they are not predetermined, they are susceptible to modifications. In other words, future presents discussed by future influencers on Twitter are conceptualized suggestions that might change if certain conditions change, but they possess significant influence in compelling others to anticipate various possibilities and take immediate action to prepare for them.

As we have shown, utilizing social data extracted from Twitter allows for the modelling of futures presented to the audience through tweets, which holds significance for several reasons. Firstly, patterns and trends can be identified, which can potentially offer predictive insights (e.g., market trends, emerging opportunities). Secondly, by anticipating future directions, people can proactively adapt and prepare for upcoming changes. Furthermore, analysing the futures presented in tweets and comments (not investigated here) will help in gaining insights into how the audience perceives and interprets these future scenarios (the audience's sentiment to certain futures). In the future, our focus will be directed towards enhancing and refining the modelling methods. Furthermore, we aim to extend the BERT model capabilities to encompass the prediction of emerging opportunities. However, we are aware that the success of using BERT for this depends on the quality and representativeness of the training data, as well as the ability of the model to capture and generalize patterns effectively.

\newpage

\bibliographystyle{ACM-Reference-Format}
\bibliography{citations}

\appendix


\end{document}